\documentclass[12pt]{article}

\usepackage{epsf}
\usepackage[dvips]{graphicx}
\begin{document}

\begin{center}
{\bfseries JETS IN $p-p$ COLLISIONS AT RHIC AND MONTE CARLO STUDY\\
 OF $z$-SCALING}

\vskip 5mm

T.G.~Dedovich$^{1,2 \sharp}$, \ M.V.~Tokarev$^{1}$

\vskip 5mm

{\small (1) {\it Joint Institute for Nuclear Research, Dubna, Russia}
\\
(2) {\it Moscow State Institute of Radio-Engineering,\\
Electronics and Automation, Moscow, Russia}
\\
$^{\sharp}${\it E-mail: dedovich@sunhe.jinr.ru }}
\end{center}

\vskip 5mm

\begin{center}
\begin{minipage}{150mm}
\centerline{\bf Abstract}
Impact of the cone algorithm parameters $E_{cut}, E_{seed}, R$
on the efficiency and characteristics of the
reconstructed jets in $p-p $ collisions  at the energy
$ \sqrt s $ = 200~GeV is studied.
The PYTHIA Monte Carlo generator is used for
event generation. The dependence of dijet production fraction on
the parton transverse momentum $\hat {p}_{\bot}$  at different
algorithm parameters is analyzed. The dependence of reconstruction
efficiency of parton energy in dijet events and two leading jets
in $N$-jet ($N_{Jet}\geq  2$) events on $E_{cut}, E_{seed}, R$
is studied.
Monte Carlo results  are compared with
predictions made in the framework of $z$-scaling and experimental
data obtained at RHIC. The independence of the slope parameter $\beta$
 of the scaling function, $\psi(z)\sim z^{-\beta}$, on the algorithm parameters
 over the energy range $E_T^{Jet}=25-60$~GeV is found. The strong dependence of
the invariant cross section and the slope parameter on the algorithm
parameters with decreasing of $E_T^{Jet}$ for $E_T^{Jet}<25$~GeV
is observed.
\end{minipage}
\end{center}

\vskip 10mm

\section{Jet-finding algorithm and jet properties}

 Perturbative Quantum Chromodynamics (QCD) predicts
 the production cross sections for parton-parton scattering
 in $p-p$ collisions at high $p_T$. The
outgoing partons from the parton-parton scattering hadronize to
form jets of particles. Calculations of high-$p_T$ jet production
involve the folding of parton scattering cross sections with
experimentally determined parton distribution functions (PDFs).
Measurements of the inclusive jet cross section, the dijet
angular distribution, and the dijet mass spectrum, are usually
used to test predictions of perturbative QCD.

We used the Monte Carlo code PYTHIA 5.7 for event generation.
The mechanism of jet production involves the hard scattering process,
 the initial and
the final state radiation and the multiple parton interaction
\cite{PYTHIA}.
The dependence of the jet reconstruction
efficiency, the parton transfers momentum $P_T^{Part}$ accuracy
and the inclusive cross sections of jet production
on the algorithm parameters are
studied. The $z$-scaling predictions for transverse jet spectra
are compared with MC-results and RHIC data.

{\subsection{Jet-finding algorithm}

 We used the adapted by D0 experiment algorithm for jet
reconstruction \cite{Snowmass}.
The jet-finding algorithm consists of the following steps:\\
1. The particles with energy $E_T^i> E_{seed}$ are assigned  as "seeds".\\
2. Start with highest-$P_T$ "seed" and iterate over all "seeds".
The "seed"-direction gives the first approximation
of the jet direction $(\eta_{Jet}, \phi_{Jet})$.\\
3. All particles with distance $R_i$ to the jet axis in the $(\eta,\phi)$
space lesser than $R$ are included into jet
\begin{equation}
R_i = \sqrt{ (\eta_{Jet}-\eta_{i})^2+ (\phi_{Jet}-\phi_{i})^2 } <
R \label{eq:r1},
\end{equation}
where $(\eta_{i}, \phi_{i})$ are the particle direction.\\
4.  The energy $E_T^{Jet}$ and the direction of jet are calculated
using (\ref{eq:r2}):
\begin{equation}
E_T^{Jet} = \sum_i E_T^i,~~  \phi_{Jet} = \sum_i E_T^i \phi_i /
\sum_i E_T^i, ~~  \eta_{Jet} = \sum_i E_T^i \eta_i / \sum_i E_T^i
\label{eq:r2}
\end{equation}
5. The steps (3)-(4) are iterated until the jet-direction is
   stable.\\
6. If $E_T^{Jet} < E_{cut}$, the jet is discarded.\\
7. Overlapping jets are merged or splitted  depending
on the energy fraction in the overlapping region.\\
8. Repeat these steps for all "seeds".\\
9. The jet directions are recalculated using an alternative
   definition as given in (\ref{eq:r4})
\begin{equation}
\theta_{Jet} = \tan^{-1} \left[ \frac{\sqrt{(\sum_i E_x^i)^2 +
(\sum_i E_y^i)^2)}}{\sum_i E_z^i}\right], ~\phi_{Jet} =
\tan^{-1}\left[ \frac{\sum_i E_y^i}{\sum_i E_x^i}\right],~
\eta_{Jet} = -\ln[\tan(\theta_{Jet}/2)] \label{eq:r4}
\end{equation}
where $i$ corresponds to all particles whose centers are within the
jet radius $R$, $E_x^i=E_i\sin\theta_i\cos\phi_i,
E_y^i=E_i\sin\theta_i\sin\phi_i, E_z^i=E_i\cos\theta_i$.

In the paper we consider $E_{seed}, R, E_{cut}$ as the parameters
of the jet-finding algorithm and analyze their influence
on characteristics of reconstructed jets.

 {\subsection{Jet reconstruction efficiency}}

We define the jet reconstruction efficiency as a probability to find two
or three jets in event.
Figure \ref{fig:Njet} shows the probability
to find two or three jets as a function  of the transfers momentum ${\hat p}_{\bot}$
of the hard process for different sets of parameters.
The left frame is for the fixed $E_{seed}, R$ and for different values of $E_{cut}$,
and the central frame is for the fixed $E_{seed}, E_{cut}$ and for different
values of $R$.
One can see that values of the algorithm parameters
$E_{cut}$, $R$ define the low limit $P_T^{lim}$
of the ${\hat p}_{\bot}$ spectrum
and the ground boundary $P_T^{sat}$ of saturation region
 for dijet events.
 For example, the dijet reconstruction probability drops very fast for
${\hat p}_{\bot}<13$~GeV  at $E_{cut}$ =7~GeV.
The low limit of the ${\hat p}_{\bot}$ spectrum increases with
$E_{cut}$ and as $R$ decreases.
The probability to find two jets for given ${\hat p}_{\bot}$
has a maximum at $E_{cut}\approx {\hat p}_{\bot}/2$.
It diminishes with decreasing of $E_{cut}$.
Part of events are reconstructed as three-jets. The right frame
is for fixed $E_{cut}$ and $R$ and different values of $E_{seed}$.
As seen from Figure \ref{fig:Njet} the influence of $E_{seed}$
on jet reconstruction is very small while $E_{seed}<<{\hat p}_{\bot}$.
 Therefore, following results are presented for  $E_{seed}$=1~GeV.
The parameters $E_{cut}$ and $R$ have an influence
on the number of reconstructed jets.
Therefore the following results are presented for dijet events and
for two leading jets of $N$-jet ($N_{Jet}> 1$).

\begin{figure}[h]
\epsfysize=50mm
\centerline{
\epsfbox{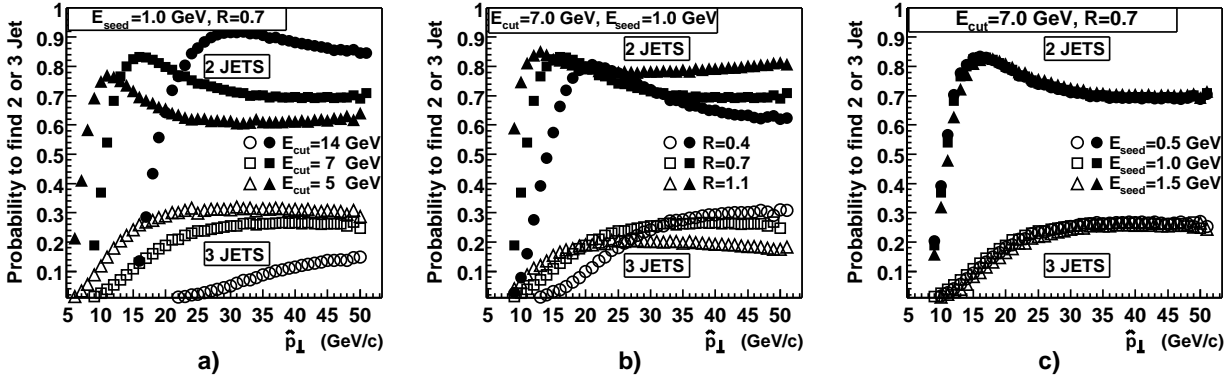}}
\caption{
The dependence of the probability to find two or three jets in $p-p$ collisions
on the transfers energy of the hard process ${\hat p}_{\bot}$
at the different values of the algorithm parameters: (a) $E_{seed}$=1.0~GeV, $R$=0.7,
$E_{cut}$=5, 7, 14~GeV; (b)  $E_{cut}$=7.0~GeV,
$E_{seed}$=1.0~GeV, $R$=0.4, 0.7, 1.1; (c)  $E_{cut}$=7.0~GeV,
$R$=0.7, $E_{seed}$=0.5, 1.0, 1.5~GeV.}
\label{fig:Njet}
\end{figure}

 {\subsection{Parton transverse momentum reconstruction}}

   We characterize the parton transverse momentum $P_{T}^{Part}$
reconstruction efficiency via the dependence of the mean jet transverse
energy $<E_{T}^{Jet}>$ and the $E_T^{Jet}$ distribution width ($RMS$) on
parton transverse momentum $P_{T}^{Part}$. These dependencies are
obtained from analysis of the dijet distribution events
as a function of jet energy $E_{T}^{Jet}$ for narrow parton
momentum bins.
One of them is presented in Figure \ref{fig:PTEJ}(a).
Figure \ref{fig:PTEJ}(b) shows the dependence
of the mean jet transverse energy as a function on $P_{T}^{Part}$ for
a fixed value of $E_{cut}$ and $E_{seed}$ and different values of  $R$.
The linear dependence of $<E_{T}^{Jet}>$ on parton transverse momentum
is observed at $P_{T}^{Part} > P_T^{lim}$.
Similar results are found for the another sets
of the jet finding algorithm parameters.
\begin{figure}[h]
\epsfysize=50mm
\centerline{\epsfbox{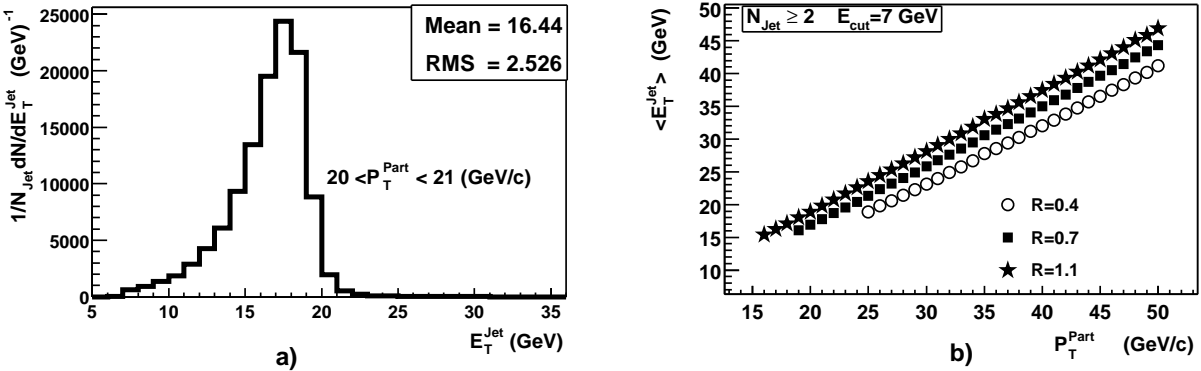}}
 \caption{(a) The dependence of number of dijet events on the jet transverse energy
$E_{T}^{Jet}$ for $20<P_{T}^{Part}<21$ (GeV/c). (b) The dependence
of the mean jet transverse energy $<E_{T}^{Jet}>$ on the parton
transverse energy $P_{T}^{Part}$ for two leading jets in $N$-jets
events ($N_{Jet} \geq 2$) at $E_{cut}$=7~GeV and $R=0.4, 0.7, 1.1$.}
 \label{fig:PTEJ}
\end{figure}

\begin{figure}[h]
\epsfysize=50mm
\centerline{
\epsfbox{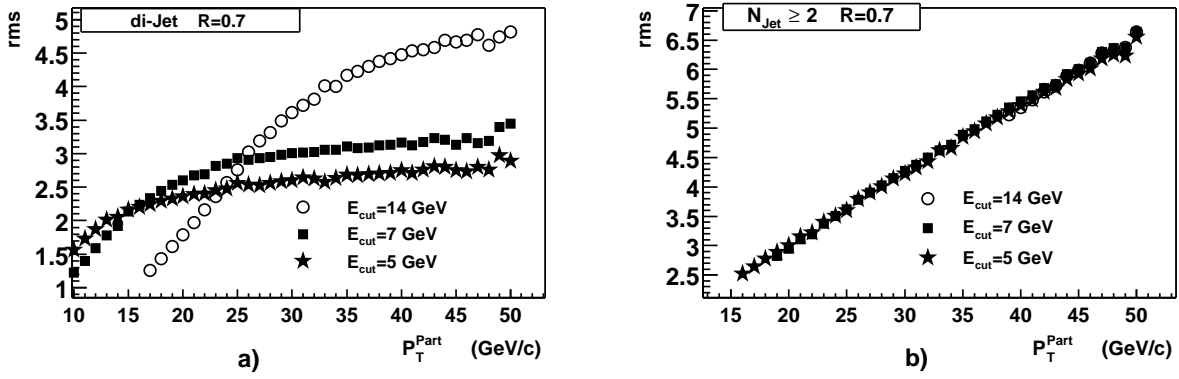}}
\caption{The dependence of $RMS$ for dijet events (a) and two leading jets in
$N$-jets events ($N_{Jet} \geq 2$) (b) on the parton transverse
momentum $P_{T}^{Part}$ at $R=0.7$ and $E_{cut}$ = 14, 7, 5~GeV.}
\label{fig:PTE}
\end{figure}

Figure \ref{fig:PTE} shows the dependence of $RMS$ as a function
of $P_{T}^{Part}$ for fixed values  of  $R$ and $E_{seed}$ and different
values of  $E_{cut}$. The left frame is for dijet events.
The slight growth of width is observed for $P_T^{Part} > P_T^{sat}$.
We would like to note, that the momentum
$P_T^{sat}$ defines the low boundary of the saturation region of
the probability to find dijet events.
The diminution of the dijet
reconstruction probability for $P_T^{Part} < P_T^{sat}$ leads to
decrease of $RMS$ values.
 The lower $E_{cut}$
provides the smaller $E_T^{Jet}$ distribution width. The dijet
reconstruction probability decreases in this case. The right frame
is for two leading jet of $N$-jet events, where $N_{Jet} > 1$.
The jet transverse energy
 distribution width is proportional to $P_T^{Part}$ in
this case. The parton transverse momentum reconstruction efficiency
is almost the same for different values of  $E_{cut}$.

\begin{figure}[h]
\epsfysize=50mm
\centerline{
\epsfbox{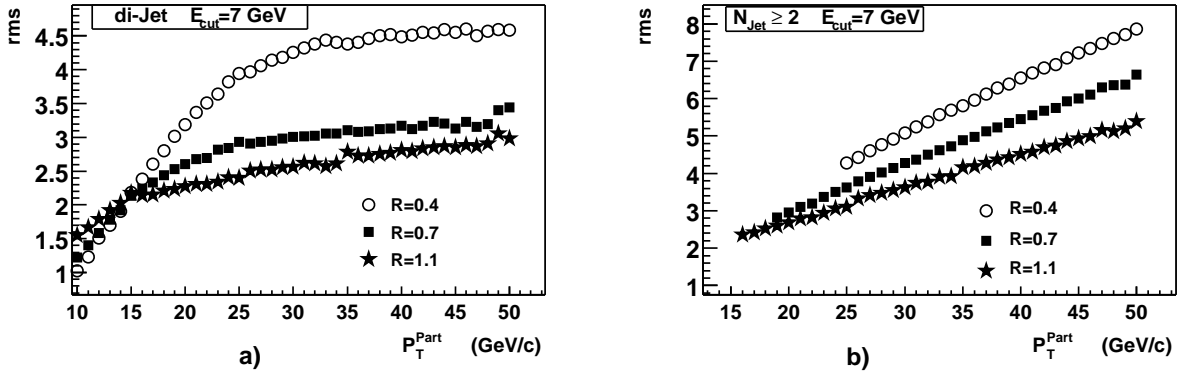}}
\caption{The dependence of $RMS$ for dijet events (a) and two leading jets in
$N$-jets events ($N_{Jet} \geq 2$)(b) on the parton transverse
momentum $P_{T}^{Part}$ at $E_{cut}$=7~GeV and $R=0.4, 0.7, 1.1$.}
\label{fig:PTR}
\end{figure}

Figure \ref{fig:PTR} demonstrates the dependence of $RMS$ on
 $P_{T}^{Part}$ for fixed values of  $E_{cut}$ and $E_{seed}$
and different values of $R$. The left frame is for dijet events.
The growth of width for $P_T^{Part} > P_T^{sat}$ and alteration of
$RMS$ for $P_T^{Part} < P_T^{sat}$ are found. The higher $R$
provides the smaller $E_T^{Jet}$ distribution width. The
difference of $RMS$ for $R=0.7$ and $R=1.1$ is found to be not so
large. Results of analysis with higher $R$ are potentially more
sensitive to background.
  The value of the $RMS$ raises substantially for $R$=0.4.
The right frame is for two leading jet of $N$-jet events ($N_{Jet} > 1$).
 The jet transverse energy distribution width is proportional
to $P_T^{Part}$. The  reconstruction efficiency is different for
various $R$.

{\subsection{Inclusive cross section of jet production}

We study the dependence of jet cross section on
the jet-finding algorithm parameters  $E_{seed}, R, E_{cut}$.
Figure \ref{fig:SIGd123} presented
 the dependence of inclusive jet cross section on the jet
transverse energy $E_{T}^{Jet}$ for one, two, three and all-jet
events (the left frame). The dependence of the cross section ratio
on $E_{T}^{Jet}$ is shown on the right frame.
One see that events with one
reconstructed jet dominate for $E_T^{Jet}<$15~GeV.
Their contributions are noticeable for  $E_T^{Jet}<25$~GeV. The dijet
events dominate for $E_T^{Jet}>15$~GeV.
The fraction of events with three
reconstructed jets is noticeable over a range 20$< E_T^{Jet}<$50~GeV.
Maximum contribution of such events is about 15\%. Behavior of jet
multiplicity versus $E_T^{Jet}$ is almost the same for another set of
parameters ($E_{cut}= 5~GeV , E_{seed}=0.5~GeV, R=0.4$). Maximum
contribution of three-jet events is $\simeq 20\% $.

\begin{figure}[h]
\epsfysize=50mm
\centerline{ \epsfbox{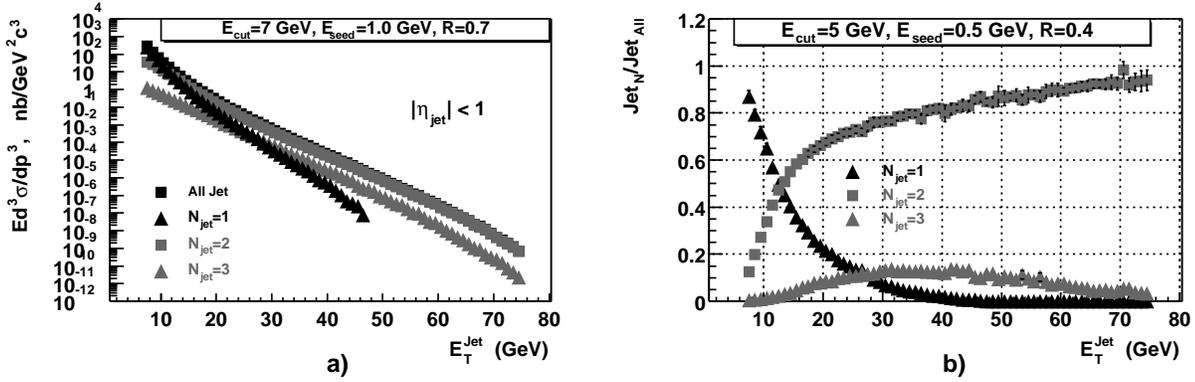}}
\caption{The dependence of the inclusive cross section (a) and the cross
section ratio $Jet_{N}/Jet_{All}$ (b) for all and $N$-jet events
($N=1, 2, 3$) on the jet transverse energy $E_{T}^{Jet}$ for PAR1
set of parameters in $p-p$ interaction at $\sqrt{s}$=200~ÃýÂ.
$(PAR1:~ E_T^{Jet}$ = 7~ÃýÂ, $E_{seed}$ = 1.0~ÃýÂ, $R=0.7$).}
\label{fig:SIGd123}
\end{figure}

\begin{figure}[h]
\epsfysize=50mm
\centerline{ \epsfbox{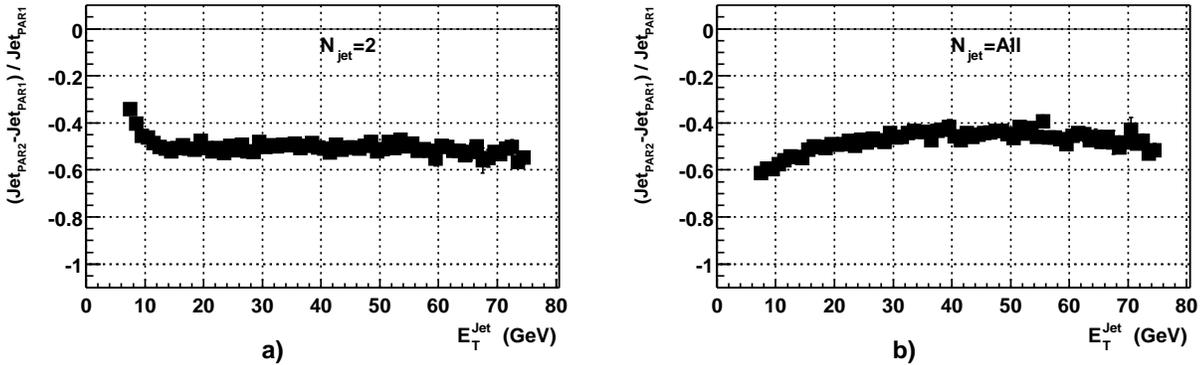}}
\caption{The dependence of the
$(Jet_{PAR2}-Jet_{PAR1})/Jet_{PAR1}$  ratio for two sets of parameters
PAR1 è PAR2 on the jet transverse energy $E_{T}^{Jet}$ for one-jet
(a) and $N$-jets ($N_{Jet}=1,2,3)$ (b) events. (PAR1: $E_T^{Jet}$
= 7~ÃýÂ, $E_{seed}$ = 1.0~ÃýÂ, $R=0.7$).} \label{fig:SIGd75}
\end{figure}

Figure \ref{fig:SIGd75} shows the dependence of
the $(Jet_{PAR2}-Jet_{PAR1})/Jet_{PAR1}$ ratio for
two sets of parameters PAR1 è PAR2 on the jet transverse energy
$E_{T}^{Jet}$. The left frame is for one-jet events. The right
frame is for $N$-jets ($N_{Jet}=1,2,3)$. As seen from Figure
\ref{fig:SIGd75} the ratio  is
practically constant above the threshold $E_T^{Sh}>15$~GeV for dijet
events. The absolute value of the obtained jet cross sections depend
on the algorithm parameters. But the energy dependence of cross
sections is practically the same for any sets of parameters above
$E_T^{Sh}=(2-3)E_{cut}$. Below this threshold the shape of cross
sections depend very strongly on parameter choice. The value of
the threshold for inclusive jet cross sections is found to be
$E_T^{Sh}=(3-5)E_{cut}$.

 Thus, above the threshold $E_T^{Sh}$
 the parameter choice results in the cross section normalization only.
 The value of cross section depends very strongly on the algorithm
 parameters below threshold.

{\section{$z$-Presentation of jet cross section}}

In this section we would like to remind the basic ideas of
$z$-scaling \cite{Z-scaling,Z-scaling2} dealing with the
investigation of the inclusive process
\begin{equation}
\ P_1 + P_2 \rightarrow p + X.\label{eq:inc}
\end{equation}
Here the momenta and masses of the colliding and inclusive particles
are denoted by $P_1, P_2, p$ and $M_1, M_2, m_1$, respectively.
It is assumed that the process can be described in terms
of the corresponding kinematic characteristics of the exclusive
subprocess written in the symbolic form
\begin{equation}
\ (x_1M_1) + (x_2M_2) \rightarrow m_1 + (x_1M_1 + x_2M_2 + m_2).
\label{eq:exc}
\end{equation}
The parameter $m_2$ is introduced to take into account
the internal conservation laws (for isospin, baryon number and
strangeness). The quantities $x_1$ and $x_2$ are
the fractions of incoming four-momenta $P_1, P_2$ of colliding
objects. The energy of the parton subprocess is defined as
\begin{equation}
\ {\hat s}^{1/2} =  {(x_1^2M_1^2 +2x_1x_2(P_1P_2)+x_2^2M_2^2)^{1/2}}.
\label{eq:esub}
\end{equation}
The elementary parton-parton collision is considered as a binary
subprocess which satisfies the 4-momentum conservation
law written in the following form
\begin{equation}
\ (x_1P_1 + x_2P_2 - p)^2 = (x_1M_1 + x_2M_2 + m_2)^2.
\label{eq:econ}
\end{equation}
To determine the fractions $x_1$ and $x_2$ the principle
of minimum resolution is used.
It states that the measure of the constituent interaction $\Omega^{-1}$
which connects kinematic and dynamic characteristics of the
interaction
\begin{equation}
\ \Omega^{-1}(x_1,x_2) = [(1-x_1)^{\delta_1}(1-x_2)^{\delta_2}]^{-1},
\label{eq:omega}
\end{equation}
should be minimum one under the condition (\ref{eq:econ}).
Here $\delta_1 $ and $\delta_2$ are the
factors relating the fractal structure of colliding objects
\cite{Z-scaling,Z-scaling2}.

The scaling variable $z$ and scaling function $\psi(z)$
as suggested in \cite{Z-scaling,Z-scaling2} are determined as follows
\begin{equation}
\ z=\frac{\sqrt{\hat s}_{\bot}}{m\rho(s)\Omega}= z_0 \Omega^{-1}, \ \ \  \psi(z) =
-\frac{\pi
s}{\rho(s,\eta)\sigma_{inl}}J^{-1}E\frac{d^3\sigma}{dp^3}
\label{eq:zpsi}.
\end{equation}
Here $m $ is the mass constant (nucleon mass), $\sqrt{\hat s}_{\bot}$
is the transverse energy of parton subprocess.

The scaling variable $z$ is expressed via the transverse  kinetic
energy of subprocess ${\hat s}_{\bot}$, the multiplicity density
of charge particles $\rho(s)$ and $\Omega(x_1,x_2)$.
The scaling function $\psi(z)$ is
expressed via the colliding energy $\sqrt s $, the average charged
particle multiplicity density $\rho(s,\eta)=d<N>/d\eta$, the inelastic
cross section $\sigma_{inl}$, the inclusive cross section
$Ed^3\sigma/dp^3$, and the Jacobian $J$ of transformation from the variables
$\{p_z,p_T\}$ to $\{\eta, z\}$.
The function $\psi$ is interpreted as the probability density
to produce a particle with the formation length $z$.
Authors of $z$-scaling concept argue that the $z$-scaling reflects
features of particle substructure, constituent interaction and
particle formation such as locality, self-similarity and fractality
over a wide scale range.

The properties of the scaling function $\psi$ of jet production
at ISR, Sp\={p}S, and Tevatron were investigated in \cite{Z-scaling,Z-scaling2}.
The energy and angular independence
of $\psi$  and power behavior, $\psi \sim z^{-\beta}$,
of jet and dijet production at high energy were established.
It was found that the values of the
slope parameter $\beta$ in $p-p$ and $\bar p-p $ collisions are different.
The properties of the scaling function $\psi$ were used to predict the
transverse energy spectrum of jet production in $p-p$ and $\bar p-p$ collisions
at RHIC, Tevatron and LHC energies.

The STAR collaboration presented \cite{Miller} the first results
on inclusive cross sections of jet production in $p-p$ collisions at RHIC.
The transverse spectra are measured over the range $\sqrt s = 200$~GeV,
$0.2<|\eta|<0.8$, and $5<E_T^{Jet}<50$~GeV.

In the present paper we analyze MC-results and  RHIC data \cite{Miller}
in $z$-presentation. The comparison with predictions of $z$-scaling is given as well.
\begin{figure}[h]
\epsfysize=50mm
\centerline{ \epsfbox{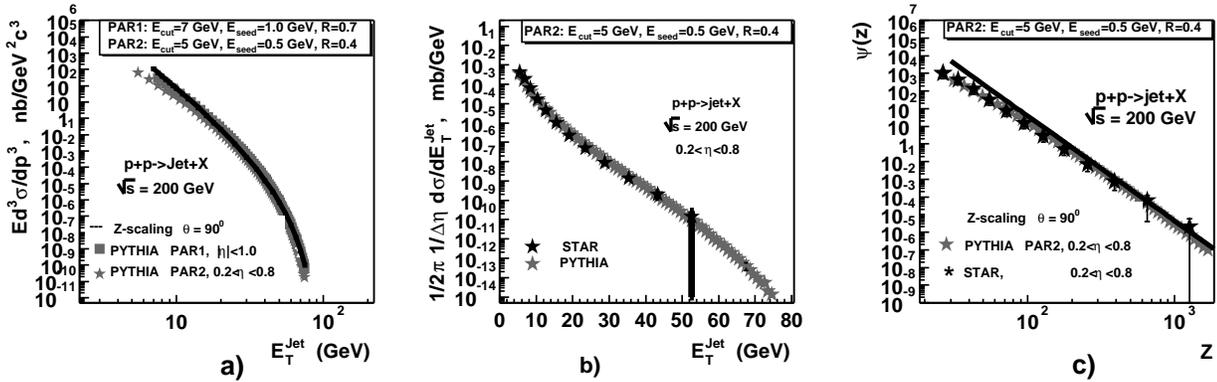}}
 \caption{The inclusive cross section of jet production in $p-p$ collisions
  at $\sqrt{s}$ = 200~ÃýÂ
in $E_T$ (a,b) and $z$ (c) presentation.}\label{fig:SIGdZ}
\end{figure}

Figure \ref{fig:SIGdZ} (a) shows Monte Carlo results
on the invariant inclusive jet cross section for sets of algorithm parameters PAR1
and PAR2. Predictions of $z$-scaling are shown by the solid line.
We found that the shape of MC jet cross sections
for different sets of parameters coincides with
 the curve predicted by $z$-scaling over a range $25<E_T^{Jet} <55$~GeV.
As seen from \ref{fig:SIGdZ} (b) the STAR date are in agreement with MC results.
The strong dependence of the cross section
shape on the algorithm parameters with decreasing $E_T^{Jet}$
is observed for $E_T^{Jet} <25$~GeV.

The comparison of the MC results, STAR data and predictions of $z$-scaling
in $z$-presentation are shown in \ref{fig:SIGdZ}(c).
We see that both MC simulation and STAR data are in a good agreement
with z-scaling predictions for $E_T^{Jet}>25$~GeV $(z>180)$.
The value of the slope parameter $\beta $ is found to be $\beta =6.01 \pm 0.06$ for
Monte Carlo results at $\sqrt s = 200$~GeV. It is  compatible with $\beta=5.95 \pm
0.21$ obtained for the AFS data at $\sqrt s = 38.8, 45$,  and 63~GeV \cite{Z-scaling2}.
Note that the shape of the scaling function $\psi(z)$
for $p_T< 10$~GeV/c can not be described by the power law $\psi(z)\sim z^{-\beta}$.
The error bars of the experimental data \cite{Miller} are large enough  for precise test of
asymptotic behavior of $z$-scaling of jet production in $p-p$ collisions at RHIC energies.

{\section{Conclusions}}

The Monte Carlo study of jet production in $p-p$ collisions
at $\sqrt s = 200$~GeV using the PYTHIA  generator was performed.
The impact of the ñone algorithm parameters
$E_{cut}, E_{seed}, R$ on the efficiency and characteristics of the
reconstructed jets was investigated.
It was established that the $P_T^{Part}$ reconstruction
efficiency is independent of $E_{cut}$ at $R=0.7$.
and depends on $R$ for two leading jets in $N$-jet events.

The probability of two-jet reconstruction was found to drop
very fast if the transverse energy of
hard process is less them the threshold ${\hat p}_{\bot} < P_T^{lim}$.
It goes to saturation if ${\hat p}_{\bot} > P_T^{sut}$.
These limits are controlled  by the values of the algorithm parameters
 $E_{seed}, R, E_{cut}$.
Results of Monte Carlo study were compared with
predictions made in the framework of $z$-scaling and experimental
data obtained by the STAR collaboration at RHIC.
The independence of the slope parameter $\beta $ of the
scaling function, $\psi(z)~z^{-\beta}$, on the algorithm parameters
 over a range  $E_T^{Jet}=25-60$~GeV was established.
 The strong dependence of
the transverse energy spectra of jet production
 and the slope parameter on algorithm
parameters were found for $E_T^{Jet}<25$~GeV.

Verification of $z$-scaling for jet production
in $p-p$ collisions at RHIC and comparison
with data at Tevatron and  LHC energies are of interest
for understanding the jet phenomena.

\vskip 5mm
{\bf Acknowledgments.}
The investigation has been partially supported
by the special program of the Ministry of Science
and Education of the Russian Federation,
grant RNP.2.2.2.2.6546.


\begin{thebibliography}{99}
\bibitem{PYTHIA} T.~Sjostrand et al., Computer Physics Commun. {\bf 135}, 238 (2001).

\bibitem{Snowmass}
B.~Abbott et al.
Phys. Rev. {\bf D 64}, 032003 (2001).

\bibitem{Z-scaling} M.V.~Tokarev, T.G.~Dedovich, JINR Preprint E2-99-300, Dubna 1999.\\
M.V.~Tokarev, T.G.~Dedovich, JINR Preprint E2-2004-188, Dubna,
2004.

\bibitem{Z-scaling2} M.V.~Tokarev, T.G.~Dedovich,
Int. J. Mod. Phys. {\bf A 15}, 3495 (2000).

\bibitem{Z-scaling3}
M.V.~Tokarev,  T.G.~Dedovich, Phys. At. Nucl. {\bf 68}, 404
(2005).

\bibitem{Miller} M.L.~Miller (for the STAR collaboration), hep-ex/0604001, 1 April, 2006.

\end{thebibliography}
\end{document}